\documentclass[12pt]{iopart}
\usepackage{graphicx}

\newcommand {\la} {\langle}\newcommand {\ra} {\rangle}
\newcommand {\beq} {\begin{eqnarray}}
\newcommand {\eeqn} [1] {\label{#1} \end{eqnarray}}
\newcommand {\eol} {\nonumber \\}
\newcommand {\ve} [1] {\mbox{\boldmath $#1$}}

\begin{document}
\title{Three-body problem with velocity-dependent optical potentials: a case of $(d,p)$ reactions}
\author{N. K. Timofeyuk }
\address{Department of Physics, Faculty of Engeneering and Physical Sciences,\\ Univesity of Surrey, Guildford, GU2 7XH, UK}
\ead{N.Timofeyuk@surrey.ac.uk}
\date{August 2018}

\begin{abstract}
    The change in mass of a nucleon, arising from its interactions with other nucleons inside the target, results in velocity-dependent terms in the Schr\"odinger equation that describes nucleon scattering. It has  recently been suggested in a number of publications that introducing and fitting velocity-dependent terms improves the quality of the description of nucleon scattering data for  various nuclei. The present paper discusses velocity-dependent optical potentials in a context of a three-body problem used to account for deuteron breakup in the entrance channel   of  $(d,p)$ reactions. Such potentials form a particular class of nonlocal optical potentials which are a popular object of modern studies. It is shown here that because of a particular structure of the velocity-dependent terms the three-body problem can be formulated in two different ways. Solving this problem within an  adiabatic approximation  results in a significant difference between the two approaches  caused by contributions from the high $n$-$p$ momenta in deuteron in one of them. Solving  the  three-body problem beyond the adiabatic approximation may  remove such contributions, which  is indirectly confirmed by replacing the adiabatic approximation by  the folding Watanabe model where such contributions are suppressed.   Discussion of numerical results is carried out for the $^{40}$Ca($d,p)^{41}$Ca reaction where experimental data both on  elastic scattering in entrance and exit channels and on nucleon transfer are available. 
\end{abstract}



\maketitle

\section{Introduction} 

It has recently been pointed out that the introduction of velocity-dependent optical potentials improves the quality of the description of proton and neutron elastic scattering  \cite{Jag11,Jag12a,Jag12b,Zur13,Gha15}. The origin of the velocity-dependence in these papers is attributed to a position-dependent effective mass of the nucleon moving in a nuclear medium,  representing another source of nonlocality.
Refs. \cite{Jag11,Jag12b}  cite   a number of papers
discussing the position-dependence of effective mass and  point to a wider range of applications for velocity-dependent potentials  such as in pion-nucleon, nucleon-nucleon and electron-atom scattering as well as in describing the dynamics of electrons in semiconductors and quantum dots. For velocity-dependent effects in nuclear matter see \cite{Gre62}.

On the other hand, it is also known that a nonlocal two-body problem is equivalent to  one with a potential that contains an infinite sum of powers of kinetic energy operators arising from  the Taylor series expansion of an exponent  that contains the nucleon kinetic energy operator \cite{PB}. Thus, from a formal point of view, the  velocity dependence used in Refs \cite{Jag11,Jag12a,Jag12b,Zur13,Gha15} is just a particular case of a more general nonlocal problem, truncated to retain linear terms only. 

The present paper discusses the velocity-dependence of optical potentials in the  context of a three-body $A+n+p$ problem and inquires how they perform when used to describe one nucleon transfer in $(d,p)$ reactions which are an important tool of spectroscopic studies of atomic nuclei.  In recent years  discussions of the role of nonlocality in $(d,p)$ studies became very popular \cite{Del09,Tim13a,Tim13b,Tit14,Ros15,Tit16,Wal16,Bai16,Bai17,Gom18,Del18,Li18}. They were triggered by three-body Faddeev calculations in \cite{Del09} showing an improvement of  the differential $(d,p)$ cross sections description when nonlocality is introduced. A few years later, methods to treat deuteron breakup within an adiabatic approximation were developed \cite{Tim13a,Tim13b,Tit16,Bai17}. However, the     adiabatic approximation was shown to induce  strong sensitivity to the high $n$-$p$ momenta in the incoming deuteron  \cite{Bai16}. The sensitivity goes away when the $A+n+p$ problem is solved beyond the adiabatic approximation, which was observed both in a  leading-order continuum discretized couped channel (CDCC)  \cite{Gom18} and   exact Fadeev \cite{Del18} calculations. Extending the  CDCC beyond the leading order is a non-trivial task while its development  for  nonlocal potentials of the velocity-dependent type introduced in \cite{Jag11} could be a simpler task. The possibility of   future CDCC applications for this class of potentials partially motivates the present work.

The paper considers  $(d,p)$ reactions within the Johnson-Tandy adiabatic distorted wave approximation (ADWA) \cite{JT}. It starts by describing the two-body problem with velocity-dependent potentials   and their local-equivalents in section 2. Then section 3 discusses the three-body Schr\"odinger equation. It is shown that velocity-dependent forces could be used in two different ways determined by their two equivalent representations in the two-body channel. Adiabatic deuteron potentials are constructed in section 4 for both these models and numerical results are reported for the case of $^{40}$Ca$(d,p)^{41}$Ca reaction in section 5. Conclusions are given in section 6.

\section{Velocity-dependent optical potentials in a two-body problem}

The optical potential parameterization in \cite{Jag11} was based on the fact 
that the change in mass of the nucleon, described by an isotropic function $\rho(r)$ and arising from its interactions with other nucleons inside the target, results in a velocity-dependence in the Schr\"odinger equation that has the simplest from given by 
\beq
(T+V(r)+\frac{\hbar^2}{2\mu} [\rho(r)\nabla^2 + \nabla\rho(r) \cdot \nabla ]-E)\Psi =0,
\eeqn{varmass}
where $V(r)$ is the local optical potential that can include spin-orbit interaction. Here and everywhere below $\mu$ denotes the reduced mass of the $N+A$ system.   Dividing Eq. (\ref{varmass}) by $1-\rho(r)$   transforms it to an equivalent equation
\beq
(T + {\tilde U}(r) + \nabla F(r) \cdot \nabla -E )\Psi = 0,
\eeqn{vdSE2}
where the  new  potential ${\tilde U}$ and the velocity-dependent force  $ F $  are given by equations 
\beq
{\tilde U}(r) &=& \frac{V(r) - E\,  \rho(r)}{1-\rho(r)}, \\
F(r) &=& -\frac{\hbar^2}{2\mu} \ln (1-\rho(r)).
\eeqn{tildu}
Eq. (\ref{vdSE2}) can  be reduced to a standard form  by introducing the Perey factor $P(r)$,
\beq
\Psi(\ve{r}) = P(\ve{r})\varphi(\ve{r})
\eeqn{}
and demanding that the equation describing $\varphi$ should not contain first derivatives. Then
  $P(\ve{r})$ satisfies the first-order differential equation
\beq
\frac{\nabla P}{P}= \frac{\mu}{\hbar^2}\nabla F
\eeqn{pereyeq}
with the condition of $P \rightarrow 1$ when $r \rightarrow \infty$, while $\varphi(\ve{r})$ satisfies the local Schr\"odinger equation
\beq
(T+{\tilde U}^{\rm eff}(r) -E)\varphi(\ve{r})=0,
\eeqn{LSE}
where $ {\tilde U}^{\rm eff} = {\tilde U}+\Delta U$ and
\beq\Delta U &=& -\frac{\hbar^2}{2\mu}\frac{\nabla^2P}{P}+\frac{\nabla F \cdot \nabla P}{P}
\eol
&=& - \frac{1}{2}\nabla^2F + \frac{1}{2}\frac{\mu}{\hbar^2}(\nabla F)^2.
\eeqn{delU}
As discussed in \cite{Jag12b}, for a particular choice of $F$, given by Eq. (4),
\beq
P(r) = \frac{1}{\sqrt{1-\rho(r)}}.
\eeqn{pvm}
This gives the following correction to the local-equivalent potential: 
\beq
\Delta U(r) =  - \frac{h^2}{2\mu} \left[ \frac{1}{4}\left(\frac{\rho^{\prime}(r)}{1-\rho(r)} \right)^2 + \frac{\rho^{\prime \prime}(r)}{2(1-\rho(r))}\right].
\eeqn{DelU2b}
The above equations are written for neutron scattering. For protons, the Coulomb interaction $V_c$ should be added to Schr\"odinger equations (\ref{varmass}), (\ref{vdSE2}) and (\ref{LSE}) while   ${\tilde U}$ is now given by
\beq
{\tilde U}(r) &=& \frac{V(r) - (E-V_c(r))\,  \rho(r)}{1-\rho(r)} .
\eeqn{tildup}
It should be noted that these transformations have been
discussed in \cite{Jag11,Jag12b} and the references   thererin as well as  in \cite{Gre62} and \cite{Nag78}. 


\section{Three-body problem with velocity-dependent nucleon optical potentials}

Let us consider a three-body $A+n+p$ problem with velocity-dependent $n-A$ and $p-A$ optical potentials. According to the previous section, the two-body nucleon scattering problem can be described in two exactly equivalent ways using the Schr\"odinger equation that contains either first derivatives only or both the first and the second derivatives. Below   both cases will be considered in the context of the three-body problem.

\subsection{Case I: first derivatives only}

This case is based on the two-body interactions ${\tilde U}_{nA}$ and  ${\tilde U}_{pA}$ given by model (\ref{vdSE2}). The corresponding  Schr\"odinger equation for the three-body wave function $\Psi(\ve{R},\ve{r})$ reads
\beq
( T_3 + V_{np}(\ve{r}) &+& {\tilde U}_{nA}(\ve{r}_n)+\nabla_n{F_n(\ve{r}_n)} \cdot \nabla_n  
+V^c_{pA}(r_{p})
\eol 
 &+& {\tilde U}_{pA}(\ve{r}_p)+ \nabla_p{F_p(\ve{r}_p)} \cdot \nabla_p - E )  \Psi(\ve{R},\ve{r})=0,
\eeqn{3bSE1}
where $\ve{r}_n$ and $\ve{r}_p$ are the coordinate-vectors of neutron and proton with respect to the target $A$, while $\ve{r} = \ve{r}_n - \ve{r}_p$ and $\ve{R} = (\ve{r}_n+\ve{r}_p)/2$. Also, the $\nabla_n$ and $\nabla_p$ denote gradients with respect to variables $\ve{r}_n$ and $\ve{r}_p$ respectively. They are related to the gradients $\nabla_R$ and $\nabla_r$  in coordinates $\ve{R}$ and $\ve{r}$, respectively, as
\beq
\nabla_n = \frac{1}{2}\nabla_R +  \nabla_r , \eol
\nabla_p = \frac{1}{2}\nabla_R -  \nabla_r.
\eeqn{nablas}
It is possible to reduce  equation (\ref{3bSE1}) to a form that does not contain first derivatives of variable $\ve{R}$.  This is convenient for  expanding the  wave function $\Psi(\ve{R},\ve{r})$ over  some basis functions of variable $\ve{r}$. We will look for a solution for  $\Psi(\ve{R},\ve{r})$ in the form
\beq
\Psi(\ve{R},\ve{r}) = P_n(\ve{r}_n) P_p(\ve{r}_p) \varphi (\ve{R},\ve{r})
\eeqn{psiPP}
requiring $P_i(r_i) \rightarrow 1$ for $r_i \rightarrow \infty$. 
Substituting it into Eq. (\ref{3bSE1}) and demanding that there are no terms containing first derivatives over $R$ we obtain 
\beq
-\frac{\hbar^2}{\mu_{dA}} \nabla_R[P_nP_p] +\frac{1}{2}P_nP_p (\nabla_n F_n +\nabla_p F_p) =0,
\eeqn{demand}
where $\mu_{dA} = m_A(m_n+m_p)/(m_A
+m_n+m_p)$ and $m_i$ is the mass of nucleus $i$.
Since $\nabla_R = \nabla_n+\nabla_p$, Eq. (\ref{demand})   after division by $P_nP_p$ becomes equivalent to  
\beq
-\frac{\hbar^2}{\mu_{dA}}\left( \frac{\nabla_n P_n}{P_n}+\frac{\nabla_p P_p}{P_p} \right) +\frac{1}{2}  \left(\nabla_n F_n +\nabla_p F_p \right) =0,
\eeqn{}
which can only be achieved when 
\beq
 \frac{\nabla_N P_N}{P_N}= \frac{\mu_{dA}}{2\hbar^2} \nabla_N F_N 
\eeqn{peq}
both for neutrons ($N=n$) and protons ($N=p$). Eq. (\ref{peq}) is almost identical to (\ref{pereyeq}) except for the reduced mass, which is different by the factor of $(m_A+1)/(m_A+2)$.
With this choice of $P_n$ and $P_p$ the wave function $\varphi(\ve{R},\ve{r})$ satisfies the Schr\"odinger equation with local nucleon optical potentials, consisting of the original local ${\tilde U}_{NA}$ potentials with some Perey-factor-based corrections, and  with additional contributions that could be considered as a three-body force because they depend on the positions of both the neutron and the proton at the same time:  
\beq
[T_3 &+&  V_{np}(\ve{r}) + {\tilde U}_{nA}(\ve{r}_n)+{\tilde U}_{pA}(\ve{r}_p)+V^c_{pA}(r_{p}) +\left[\frac{\nabla_n P_n \cdot \nabla_n F_n}{P_n}    + \frac{\nabla_p P_p \cdot \nabla_p F_p}{P_p}  \right]  
\eol 
 &+& \left(1-\frac{\mu_{dA}}{4\mu_{np}}\right)( \nabla_n{F_n}-\nabla_p{F_p} )\cdot \nabla_r 
 -\left(\frac{\hbar^2}{\mu_{dA}}-\frac{\hbar^2}{4\mu_{np}}\right)\frac{\nabla_nP_n\cdot\nabla_pP_p}{P_nP_p}
 \eol
 &-&  \left( \frac{\hbar^2} {2\mu_{dA}}+\frac{\hbar^2}{8\mu_{np}} \right)  \frac{\nabla^2_nP_n}{P_n}
 -
  \left( \frac{\hbar^2} {2\mu_{dA}}+\frac{\hbar^2}{8\mu_{np}} \right) \left. \frac{\nabla^2_pP_p}{P_p} - E\right] \varphi(\ve{R},\ve{r})=0.
 \eol
\eeqn{3bSE}
One of these contributions has $n-p$ velocity-dependence that comes through $\nabla_r$.
If one assumes all reduced masses  are determined by nuclei mass numbers only then
\beq 
\frac{1} {\mu_{dA}}+\frac{1}{4\mu_{np}}=\frac{1} {\mu},
\eol
\frac{1} {\mu_{dA}}-\frac{1}{4\mu_{np}}=\frac{1} {M_A},
\eeqn{redmas}
where $\mu$ is the  reduced mass of the $A+1$ system, the same one that features in the two-body problem, and $M_A$ is the mass of the target $A$. In this case the Schr\"odinger equation reduces to
\beq
[T_3 &-& E + V_{np}(\ve{r}) + {\tilde U}^{\rm eff} _{nA}(\ve{r}_n)+{\tilde U}^{\rm eff}_{pA}(\ve{r}_p) 
+V^c_{pA}(r_{p})
\eol 
 &+& \frac{2}{A+2}( \nabla_n{F_n}-\nabla_p{F_p} )\cdot \nabla_r 
 -\left.\frac{\mu^2_{dA}}{4\hbar^2M_A}
 \nabla_nF_n\cdot\nabla_pF_p
\right] \varphi(\ve{R},\ve{r})=0
 \eol
\eeqn{3bSEIa}
where the effective $N-A$  potentials are given by
\beq
{\tilde U}^{\rm eff}_{NA}  = {\tilde U}_{NA}  -\frac{1}{2} \frac{A+1}{A+2} \nabla^2_N F_N+ \left( 1 - \frac{1}{2} \frac{A+1}{A+2} \right) \frac{\mu_{dA}}{2\hbar^2} (\nabla_NF_N)^2.
\eeqn{}
They differ from two-body effective potentials of section 2 by additional mass-dependent factors that  vanish when $A \rightarrow \infty$.

\subsection{Case II: first and second derivatives}

This case is based on the original   model (\ref{varmass}) with  local  interaction  $V_{NA}(r)$. The corresponding  Schr\"odinger equation for the three-body wave function $\Psi(\ve{R},\ve{r})$ reads
\beq
( T_3 + V_{np}(\ve{r}) &+& V_{nA}(\ve{r}_n)+\frac{\hbar^2}{2\mu}\left(\nabla_n{\rho_n(r_n)} \cdot \nabla_n  + \rho_n(r_n) \nabla_n^2 \right)
+V^c_{pA}(r_{p})
\eol 
 &+& V_{pA}(\ve{r}_p)+ \frac{\hbar^2}{2\mu}\left(\nabla_p{\rho_p(r_p)} \cdot \nabla_p  + \rho_p(r_p) \nabla_p^2\right)- E )  \Psi(\ve{R},\ve{r})=0.
 \eol
\eeqn{3bSE2}
Because of the second-order derivatives $\nabla_n^2$ and $\nabla_p^2$, the introduction of the Perey factors from the previous subsection will result in an additional term $(\nabla_R \cdot \nabla_r)$ in the Schr\"odinger equation (\ref{3bSEIa}), which does not bring any advantages. Therefore, no preliminary modifications of Eq. (\ref{3bSE2}) have been done. 

\section{Adiabatic approximation for $(d,p)$ reactions with velocity-dependent potentials}

The  $(d,p)$  transition amplitude contains the short-range interaction $V_{np}$ \cite{Satchler} and therefore the $(d,p)$ cross sections   are determined by the wave function $\Psi(\ve{R},\ve{r})$   at very small values of $r$. Usually, for local $V_{NA}$ potentials,  $\Psi(\ve{R},\ve{r})$ is expanded over the Weinberg state basis and only the first term of this expansion is retained thus assuming that $\Psi(\ve{R},\ve{r})\approx \chi(\ve{R}) \phi_0(\ve{r})$, where $\phi_0$ is the $s$-wave deuteron wave function,  is a good approximation  in the small $r$ region \cite{JT}. In this case,  $\chi(\ve{R})$ satisfies the two-body Schr\"odinger equation with the Johnson-Tandy adiabatic potential
\beq
  U^{JT}_{dA}(\ve{R}) = \int d\ve{r} \,\phi^*_1(\ve{r})\left[ V_{nA}(\ve{R}+\frac{\ve{r}}{2})+V_{pA}(\ve{R}-\frac{\ve{r}}{2})\right]  \phi_0(\ve{r}),
\eeqn{JT}
where 
\beq
\phi_1(r) = \frac{\phi_0(r)}{\la \phi_0|V_{np}|\phi_0\ra}. 
\eeqn{phi1}
In this paper, the same adiabatic approximation with the $s$-wave deuteron is used. Including the $d$-wave does not change $U^{JT}_{dA}(\ve{R})$ if the $N-A$ potentials are local \cite{Bai17}, however,  it produces dramatic changes in the case of nonlocal potentials \cite{Bai16}  which are an artefact of  the adiabatic approximation \cite{Gom18,Del18}.

In the case I of the previous section, the adiabatic approximation could be introduced either before or after introduction of the Perey factors. Both cases will be considered here, labelled by $(a)$, which  corresponds to introduction of adiabatic approximation $\it after$ introduction of Perey factor,
\beq
(a):   \,\,\,\,\,\,\,\,\,\,\,\,\,\,\,\,\,\,\,\, \varphi(\ve{R},\ve{r})=\chi(\ve{R}) \phi_0(\ve{r}),
\eeqn{AAa}
or $(b)$, which corresponds to introduction of adiabatic approximation $\it before$ the Perey factors are introduced:
\beq
(b):   \,\,\,\,\,\,\,\,\,\,\,\,\,\,\,\,\,\,\,\, \Psi(\ve{R},\ve{r})=\chi(\ve{R}) \phi_0(\ve{r}).
\eeqn{AAb}
As for  the case II, the adiabatic approximation will be introduced from the very beginning using approximation (\ref{AAb}).

\subsection{Case Ia}

Using approximation (\ref{AAa}) in Eq. (\ref{3bSEIa}) one obtains the following two-body Schr\"odinger equation for
 $\chi(\ve{R})$:
\beq
\left[T_2+{\tilde U}^{(0)}_{dA}(\ve{R})+\Delta U_1(\ve{R})+\Delta U_2(\ve{R})+V^c_{dA}(R)
-E_d\right]\chi(\ve{R})=0,
\eeqn{AE}
where  $T_2$ is the kinetic energy operator associated with variable $\ve{R}$, $E_d = E + \varepsilon$ is the centre-of-mass energy of the incoming deuteron ($\varepsilon$ being the (negative) deuteron binding energy) and
\beq
{\tilde U}^{(0)}_{dA}(\ve{R}) = \int d\ve{r} \,\phi^*_1(\ve{r}) \left[{\tilde U}^{\rm eff}_{nA}(\ve{R}+\frac{\ve{r}}{2})+{\tilde U}^{\rm eff}_{pA}(\ve{R}-\frac{\ve{r}}{2})\right] \phi_0(\ve{r})
\eeqn{adpot0}
is analogous to the commonly used Johnson-Tandy potential (\ref{JT}). In this paper, a  usual assumption is made that the Johnson-Tandy potential associated with the Coulomb $p-A$ potential results in the $d-A$ Coulomb potential without any polarization terms. Other terms in Eq. (\ref{AE}) are
\beq
\Delta U_1(\ve{R}) &=&  \frac{2}{A+2}
\int d\ve{r} \,\phi^*_1(\ve{r}) 
( \nabla_n{F_n}-\nabla_p{F_p} )\cdot \nabla_r  \phi_0(\ve{r}), \\
\Delta U_2(\ve{R}) &=& -\frac{\mu^2_{dA}}{4\hbar^2M_A} \int d\ve{r} \,\phi^*_1(\ve{r}) 
( \nabla_nF_n\cdot\nabla_pF_p ) \,\phi_0(\ve{r}).
\eeqn{deladpots1}
To estimate  these additional terms, the $F_n$ and $F_p$ are assumed to be spherically-symmetric. In this case
\beq
\nabla_iF_i = \ve{e}_i \frac{\partial F_i(r_i)}{\partial r_i}= \ve{e}_i F^{\prime}_i,
\eeqn{partials}
where $\ve{e}_i$ is the unit vector in the direction of $\ve{r}_i$. Similarly,
\beq 
\nabla_r \phi_0(\ve{r})  =\ve{e}_r \phi^{\prime}_0(r )
\eeqn{partialphi}
(note that $\phi_0(r)$ includes $Y_{00}(\hat{\ve{r}})$). Using relations
\beq
(\ve{e}_n\cdot \ve{e}_r) &=& \frac{1}{r_n}\left(R\nu + \frac{1}{2} r\right),\\
(\ve{e}_p\cdot \ve{e}_r) &=& \frac{1}{r_p}\left(R\nu - \frac{1}{2} r\right),\\
(\ve{e}_n\cdot \ve{e}_p) &=& \frac{1}{r_nr_p}\left(R^2 - \frac{1}{4} r^2\right),
\eeqn{ee}
where $\nu = \cos(\hat{\ve{R},\ve{r}})$ one obtains
\beq
\Delta U_1(R) &=& 
  \frac{2}{A+2} \int  d\ve{r} \,  \phi_1(r) \left[\frac{r}{2}\left(\frac{F^{\prime}_n(r_n)}{r_n} +\frac{F^{\prime}_p(r_p)}{r_p}\right) \right.
  \eol
  &+&\,\,\,\,\,\,\,\,\,\, \,\,\,\,\,\,\,\,\,\,\,\,\,\,\,\,\,\,\,\,\,\,\,\,\,\,\,\,\,\,\,\,\,\,\,\,\,\,\,\,R\nu \left. \left(\frac{F^{\prime}_n(r_n)}{r_n} -\frac{F^{\prime}_p(r_p)}{r_p}\right)\right] \phi^{\prime}_0(r), 
\eeqn{}
\beq 
\Delta U_2(R) &=& - \frac{\mu^2_{dA}}{4 \hbar^2 M_A} \int d\ve{r} \,\phi^*_1(\ve{r}) 
\frac{F^{\prime}_n(r_n) F^{\prime}(r_p)}{r_nr_p}  
\left( R^2 - \frac{1}{4}r^2\right) \,\phi_0(\ve{r}).
\eeqn{deladpots2}
in which $r_n = \sqrt{R^2+\nu r R + r^2/4}$ and  $r_p = \sqrt{R^2-\nu r R + r^2/4}$.

\subsection{Case Ib}

Making approximation (\ref{AAb}) in eq. (\ref{3bSE1}) one obtains for distorted wave $\chi(\ve{R})$ the equation
\beq
\left[T_2+{\tilde U}^{(0)}_{dA}(\ve{R})+\ve{S}(\ve{R})\cdot \nabla_R+D(\ve{R})+V^c_{dA}(R)
-E_d\right]\chi(\ve{R})=0,
\eeqn{AEIb}
where 
\beq
{\tilde U}^{(0)}_{dA}(\ve{R}) = \int d\ve{r} \,\phi^*_1(\ve{r}) \left[{\tilde U}_{nA}(\ve{R}+\frac{\ve{r}}{2})+{\tilde U}_{pA}(\ve{R}-\frac{\ve{r}}{2})\right] \phi_0(\ve{r}),
\eeqn{adpot0_Ib}
is  the Johnson-Tandy potential based on the original local-equivalent potentials ${\tilde U}_{NA}$, and
\beq
\ve{S}(\ve{R}) = \frac{1}{2}\int d\ve{r} \,\phi_1(\ve{r}) \left[ \nabla_n F_n + \nabla_p F_p\right]
\phi_0(\ve{r}),
\eeqn{SR}
\beq
D(\ve{R}) =  \frac{A+2}{2} \Delta U_1(\ve{R}).
\eeqn{DR}
It can be shown that for a spherically-symmetrical velocity-dependent force $F_N$ and for the $s$-wave deuteron the term $\ve{S}(\ve{R})\cdot \nabla_R$ is equal to
\beq
\ve{S}(\ve{R})\cdot \nabla_R =S(R) \frac{\partial}{\partial R}.
\eeqn{}
This can be obtained using Eqs. (\ref{partials}), (\ref{partialphi}) and relations $\ve{e}_R \cdot \nabla_R = \frac{\partial}{\partial R}$ and
\beq
\int d\ve{r} \, F(|\ve{r}-\ve{R}|)G(r) \, \ve{e}_r\cdot \nabla_R = \int_0^{\infty} dr \, r^2 G(r){\cal  F}_1(r,R) \frac{\partial}{\partial R},
\eeqn{}
where ${\cal F}_{\lambda}(r,R)$ is the radial part of the multipole expansion of $F(\ve{r},\ve{R})$,
\beq
F(|\ve{r}-\ve{R}|) = \sum_{\lambda\mu} {\cal F}_{\lambda}(r,R) Y^*_{\lambda\mu}(\hat{\ve{r}}) Y_{\lambda\mu}(\hat{\ve{R}}).
\eeqn{}
$S(R)$ is given by
\beq
S(R) = \int_0^{\infty} dr \, r^2 \phi_1(r) \left[ \frac{1}{2}{\cal F}_0(r,R) + \frac{1}{4}{\cal F}_1(r,R) \right]\phi_0(r)
\eeqn{sr}
with
\beq
{\cal F}_0 (r,R)= 2\pi R \int_{-1}^1 d\mu \, \left(\frac{F^{\prime}_n( r_n)}{r_n} + \frac{F^{\prime}_p(r_p)}{r_p}\right),\\
{\cal F}_1 (r,R)= 2\pi r \int_{-1}^1 d\mu \, \mu \left(\frac{F^{\prime}_n( r_n)}{r_n} - \frac{F^{\prime}_p(r_p)}{r_p}\right).
\eeqn{flam}
The Perey factor $P(R)$ can be now introduced in the usual way, $\chi(R)=P(R) \varphi(R)$. $P(R)$ satisfies the equation
\beq
\frac{P^{\prime}(R)}{P(R)} = \frac{\mu_{dA}}{\hbar^2}S(R),
\eeqn{PIb}
while $\varphi(R)$ is the solution of the ordinary Schr\"odinger equation
\beq
(T_2 + U^{\rm eff}_{dA}(R)+V^c_{dA}(R) - E_d )\varphi(R)=0
\eeqn{SELEIb}
with the effective potential
\beq
U^{\rm eff}(R) = {\tilde U}^{(0)}_{dA}(R)+V^c_{dA}(R) +\frac{1}{2} \frac{\mu_{dA}}{\hbar^2}S^2(R)-\frac{1}{2}S^{\prime}(R) +D(R).
\eeqn{UeffIb}

\subsection{Case II}

One can show that applying approximation (\ref{AAb}) in the three-body Schr\"odinger equation (\ref{3bSE2}) results in a two-body differential equation similar to Eq. (\ref{AEIb}) that describes the case Ib but with an additional term  that arises due to the second derivatives in the velocity-dependent force,
\beq
\frac{\hbar^2}{2\mu} \int d\ve{r} \,\phi^*_1(\ve{r}) (\rho_n \nabla^2_n + \rho_p \nabla^2_p)\phi_0(\ve{r})
\eol
= \frac{\hbar^2}{2\mu} \int d\ve{r} \,\phi^*_1(\ve{r}) \left[\frac{1}{4}(\rho_n+\rho_p) \nabla^2_R + (\rho_n+\rho_p) \nabla^2_r + (\rho_n - \rho_p) (\nabla_r\cdot\nabla_R)\right] \phi_0(\ve{r}).
\eol
\eeqn{AEII}
This two-body equation reads
\beq
\left(-\frac{\hbar^2}{2\mu_{dA}}\left(1 -  {\cal R} (R)\right)\nabla^2_R \right.
 &+& V^{JT}_{dA}(R) + V^c_{dA}(R)+B(R) +{\tilde D}(R)
 \eol 
 &+&  \left. (S_1(R)+S_2(R)) \frac{\partial}{\partial R}-E_d\right) \chi(\ve{R})=0,
 \eol
\eeqn{SEII}
where $V^{JT}_{dA}(R)$ is the usual Johnson-Tandy potential, constructed from the original local potentials $V_n$ and $V_p$, and 
\beq
{\cal R} (R) &=&\frac{\mu_{dA}}{4\mu}\int d\ve{r} \, \phi_1(\ve{r}) (\rho_n+\rho_p) \phi_0(\ve{r}),\\
B(R) &=& \frac{\hbar^2}{2\mu} \int d\ve{r} \,\phi_1(\ve{r})  (\rho_n+\rho_p) \nabla^2_r  \phi_0(\ve{r}). 
\eeqn{}
The $S_1(R)$ and  ${\tilde D}(R)$  are equivalent to $S (R)$ and  $ D(R)$ from the previous subsection and they are given by   Eqs. (\ref{SR}) and (\ref{DR}), respectively, by replacing  $F^{\prime}_N$ by $\hbar^2/(2\mu) \rho_N$. The $S_2(R)$  in (\ref{AEII}) is
\beq
S_2(R) = \int  d\ve{r}\, \nu \,\phi_1(r)  [\rho_n(r_n) - \rho_p(r_p)  \phi^{\prime}_0(r).
\eeqn{}
Again, as in sec. II, dividing  (\ref{AEII}) by $(1 - {\cal R}(R))$ results in
\beq
\left(T_2
 + {\tilde V}_{dA}(R) + {\tilde S}(R) \frac{\partial}{\partial R}+V^c_{dA}(R)-E_d\right) \chi(\ve{R})=0,
 \eol
\eeqn{SEIIb}
where
\beq
{\tilde V}_{dA}(R) &=&  \frac{V^{JT}_{dA}(R)  -[E_d - V^c_{dA}(R) ]\,{\cal R}(R) +  B(R) +{\tilde D}(R)}{1-{\cal R}(R)} ,
\eeqn{VdAR}
\beq
{\tilde S}(R) &=& \frac{ S_1(R)+{S}_2(R)}{1-{\cal R}(R)}.
\eeqn{}
As in the case Ib, the introduction of the Perey factor through $\chi(R) = P(R) \varphi(R)$ will help to get rid of the first derivatives over $R$. The  $P(R)$ and $\varphi(R)$ are given by Eq. (\ref{PIb}) and (\ref{SELEIb})-(\ref{UeffIb}) in which $S(R)$ is substituted by  ${\tilde S}(R)$.

\section{Numerical calculations for $^{40}$Ca(d,p)$^{41}$Ca reaction}

In this sections   the $^{40}$Ca(d,p)$^{41}$Ca cross sections are calculated for the deuteron incident laboratory energy of 20 MeV in the zero-range approximation. Calculations are done with the help of the latest version of the code TWOFNR \cite{twofnr} which has an option to read distorted waves in,  useful for modifying them externally by   non-standard Perey factors.

\begin{figure}[b]
\centering
\includegraphics[scale=0.25]{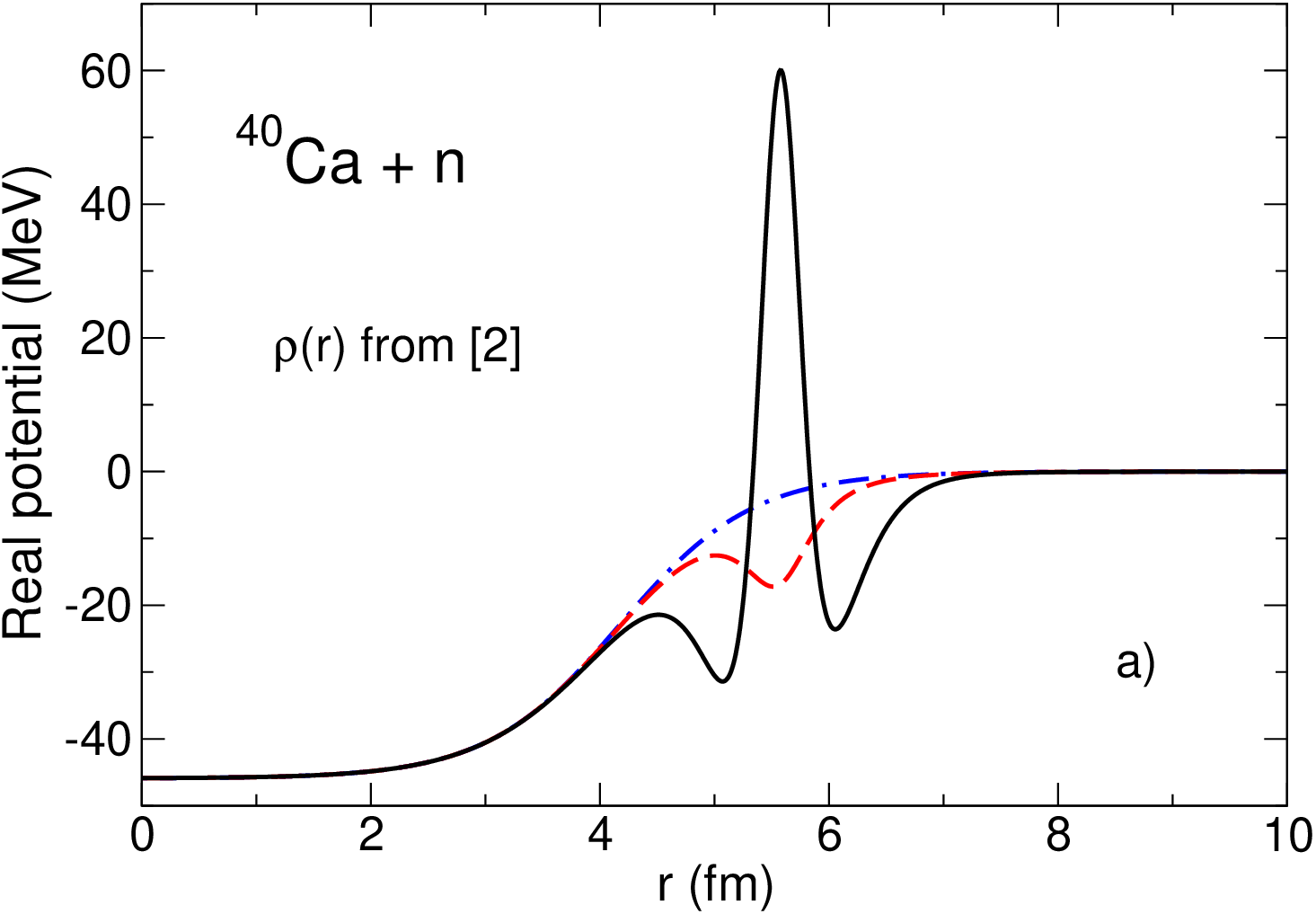}
\includegraphics[scale=0.25]{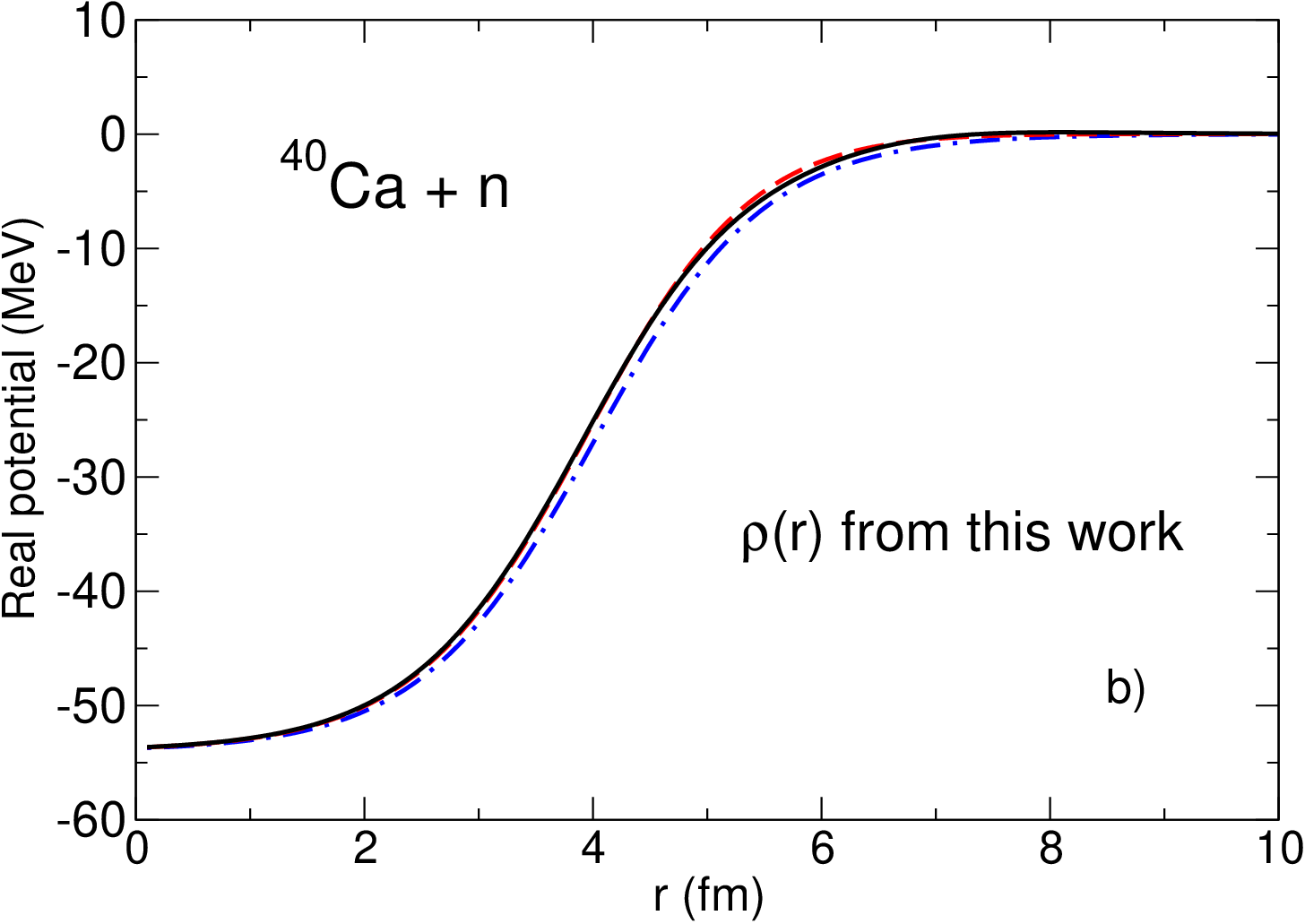}
\caption{The real parts of the $n-^{40}$Ca potentials  $V(r)$ (dot-dash line), ${\tilde U}$ (dashed line) and ${\tilde U}^{\rm eff}$ (solid line) calculated ($a$) for 11 MeV neutrons using velocity-dependent potential from \cite{Jag12a} and ($b$) for 10 MeV neutrons using Becchetti-Greenlees potential and $\rho(r)$ fitted in this work. }
\label{fig:Jagpot}
\end{figure}

The ADWA uses nucleon optical potentials taken at half the deuteron incident energy - the tradition that   goes back to works \cite{JSNP,JT}. Experimental data on neutron and proton scattering at 10 MeV and optical potentials are available. First of all,  velocity-dependent optical potential from Ref. \cite{Jag12a} was used in this paper. It is described by the function
\beq
\rho(r) = \rho_0 a_{\rho} \frac{d}{dr} \frac{1}{1+\exp [(r-r_{\rho} A^{1/3})/a_{\rho}]}
\eeqn{}
with $\rho_0 = -1.9$, $r_{\rho} = 1.63$ fm and $a_{\rho} = 0.25$ fm.
It was found that the   local equivalent  potentials  ${\tilde U}^{\rm eff}$, constructed with these parameters, have a huge repulsive narrow peak (with the height of the order of 60 MeV) in the surface region, which is clearly unphysical (see Fig. \ref{fig:Jagpot}). This peak originates due to a very narrow width of  $\rho(r)$ determined  by a small diffuseness parameter of 0.25 fm. The second derivative of this narrow function, that determines $\Delta U$ in eq. (\ref{DelU2b}), is very large. The local equivalent potential ${\tilde U}^{\rm eff}$  in the exit channel looks even more unphysical. Because of the presence of the volume term in $\rho(r)$ \cite{Zur13} at the energy of the exit proton channel, the depth of ${\tilde U}^{\rm eff}$  is only about 10 MeV, which severely affects the $(d,p)$ cross sections. The unphysical behaviour of the nucleon local equivalent potentials, based on parameters of \cite{Jag12a}, translates into unphysical behaviour of deuteron adiabatic potentials and results in a huge differences between adiabatic deuteron potentials calculated in  models Ia, Ib and II. For this reason,   the potentials of Refs. \cite{Jag12a,Zur13} were abandoned and a more conventional diffuseness was adopted for $\rho(r)$. 

\begin{figure}[h!]
\centering
\includegraphics[scale=0.3]{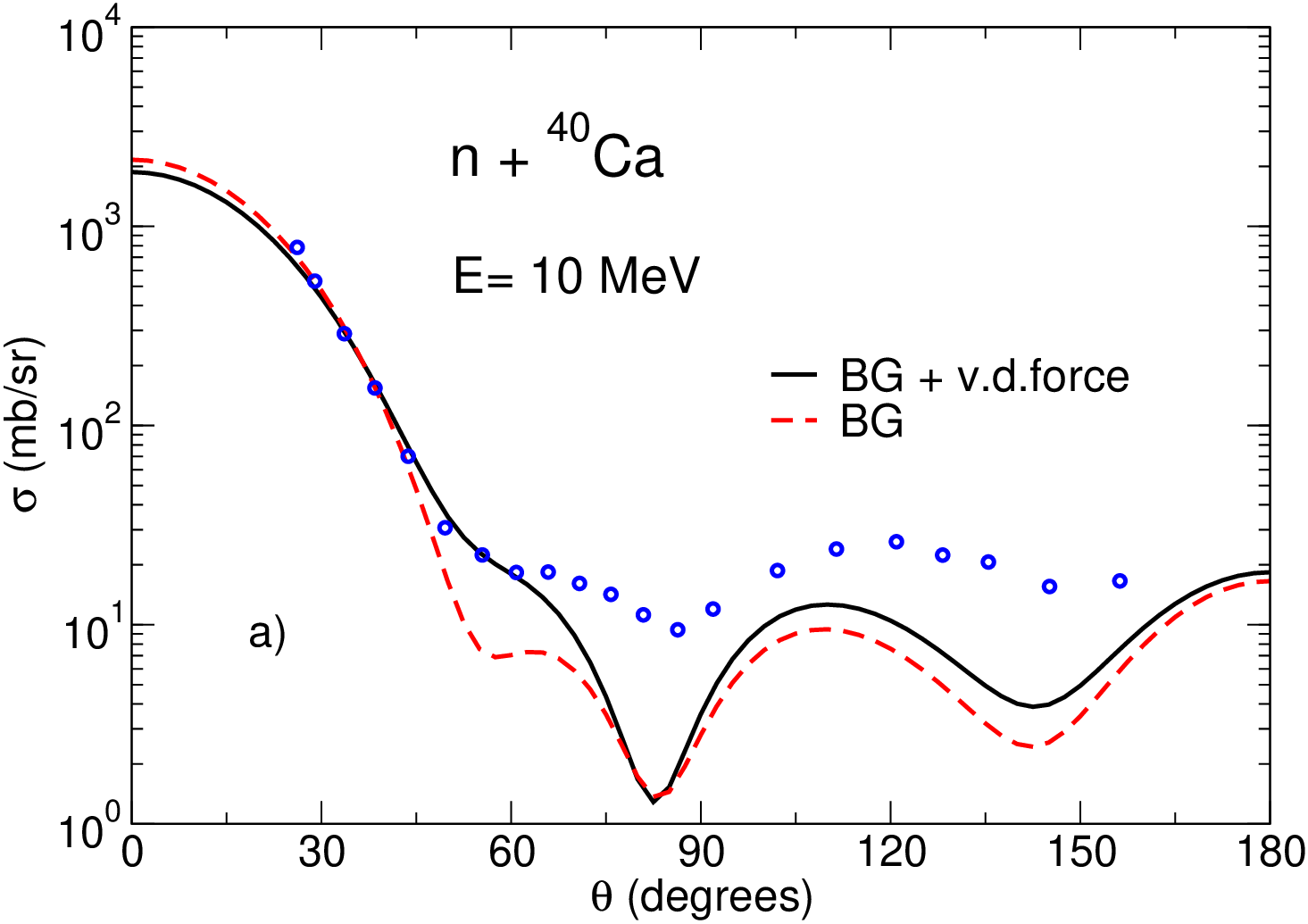}
\includegraphics[scale=0.3]{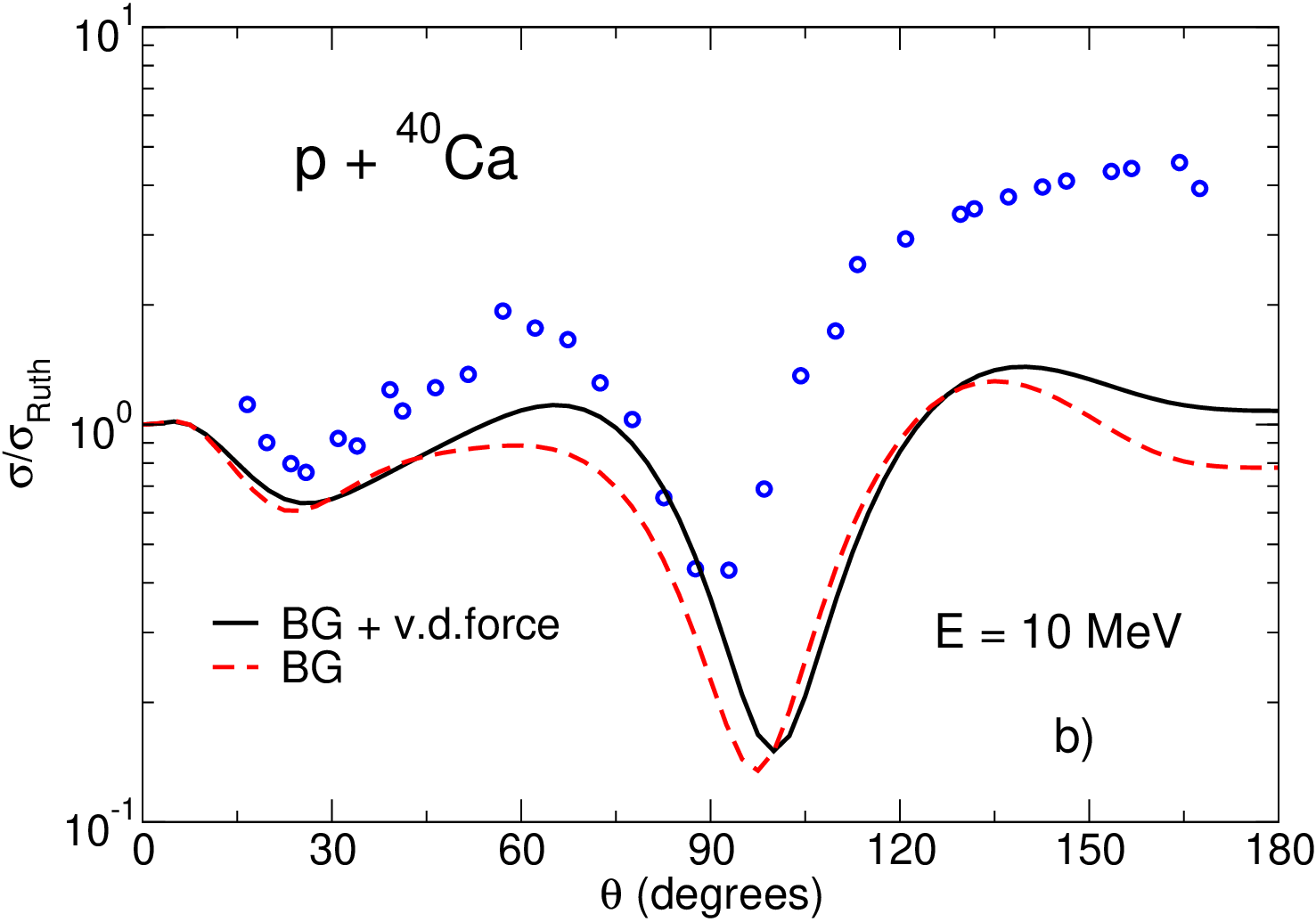}
\includegraphics[scale=0.3]{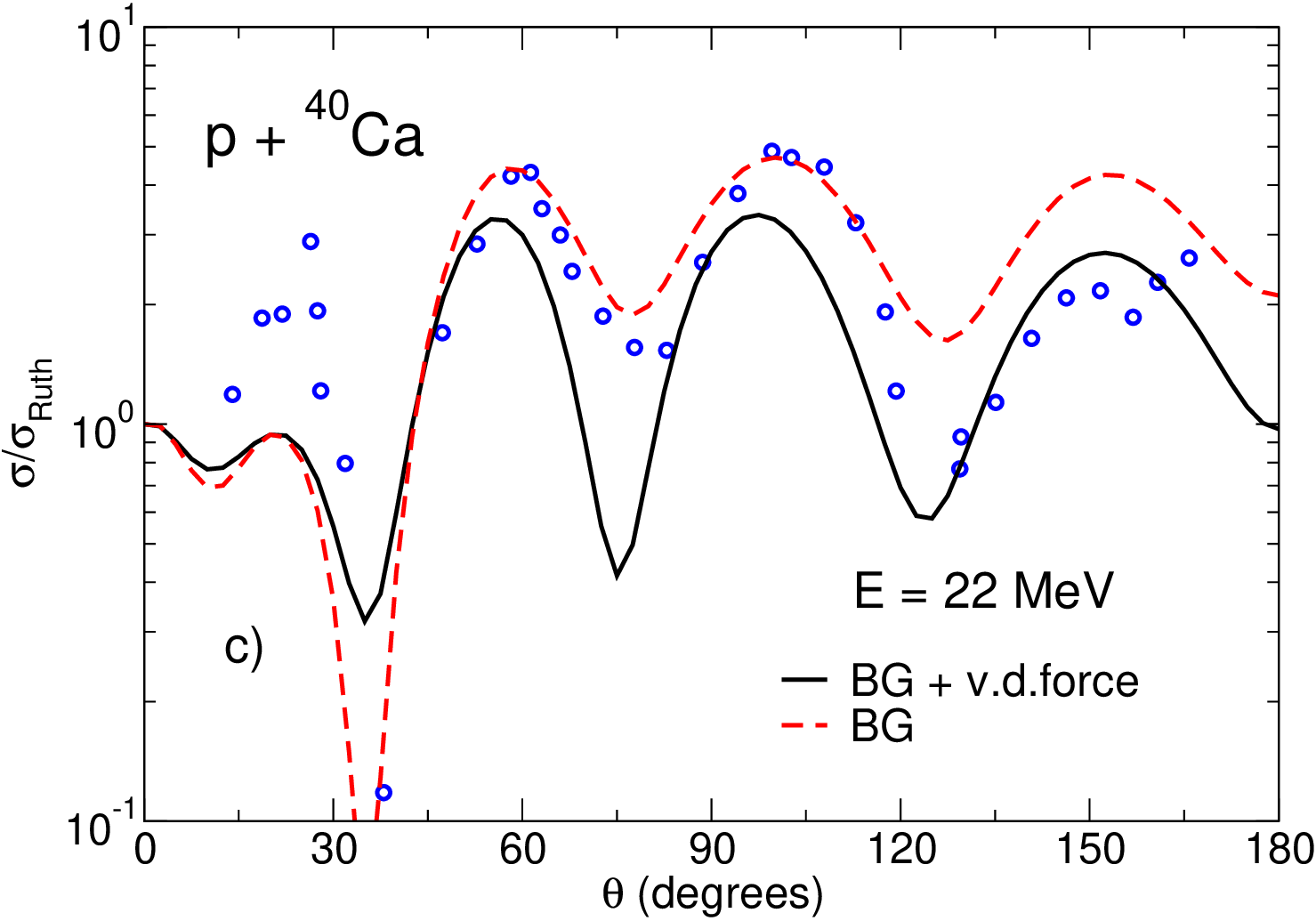}
\caption{The neutron ($a$) and proton ($b,c$) scattering from $^{40}$Ca at $E = 10 $ MeV ($a,b$) and 22 MeV ($c$) calculated using the Becchetti-Greenlees potential on its own (dashed lines) and with addition of velocity-dependent optical  potentials (solid lines) in comparison with experimental data from \cite{Tor82,Dic71}.
}
\label{fig:elastic}
\end{figure}

In all calculations below, the frequently used Becchetti-Greenlees optical potential \cite{BG} was used, to which a velocity-dependent term was added with the parameters   fitted to improve the description of the elastic scattering data from the $^{40}$Ca target  for   neutrons at 10 MeV \cite{Tor82} and for protons  at 10 and 22 MeV  \cite{Dic71}. It was found that the choice of $\rho_0= 0.4$, $r_{\rho} = 1.63$ fm and $a_{\rho} = 0.90 $ fm somewhat improves the  description of elastic scattering of neutrons and protons on $^{40}$Ca at 10 MeV (see Fig. \ref{fig:elastic}). The same set of parameters improves description of the 22 MeV data for $p+^{40}$Ca at large angles while making it a bit worse for $60 \le \theta \le 120$ degrees. The  general quality of the data fit is similar to the one with the original Becchetti-Greenless potential. These new values of $\rho_0$, $r_{\rho}$ fm and $a_{\rho}  $ are used below. It should be noted that a positive value of $\rho_0$ was essential to improvement of the data fit. Negative values deteriorated it significantly. It should also be mentioned that the surface form of $\rho(r)$, chosen in refs. \cite{Jag11,Jag12a,Jag12b}, gives a surface-peaked effective nucleon mass $m^*(r)/m = 1/(1-\rho(r))$,  equal to the square of the nucleon Perey factor. Detailed mean-field calculations of variable nucleon mass in finite nuclei  suggest that   $m^*(r)/m$ has a volume form, decreasing from  1 at large $r$    to $~0.8$ at $r=0$ \cite{Sai97}. In this paper, the effects of the surface form of $\rho(r)$ are investigated. The question of the importance of the  volume form  is left for future studies.


\begin{figure}[h!]
\centering
{
\includegraphics[width=0.70\textwidth,clip=true]{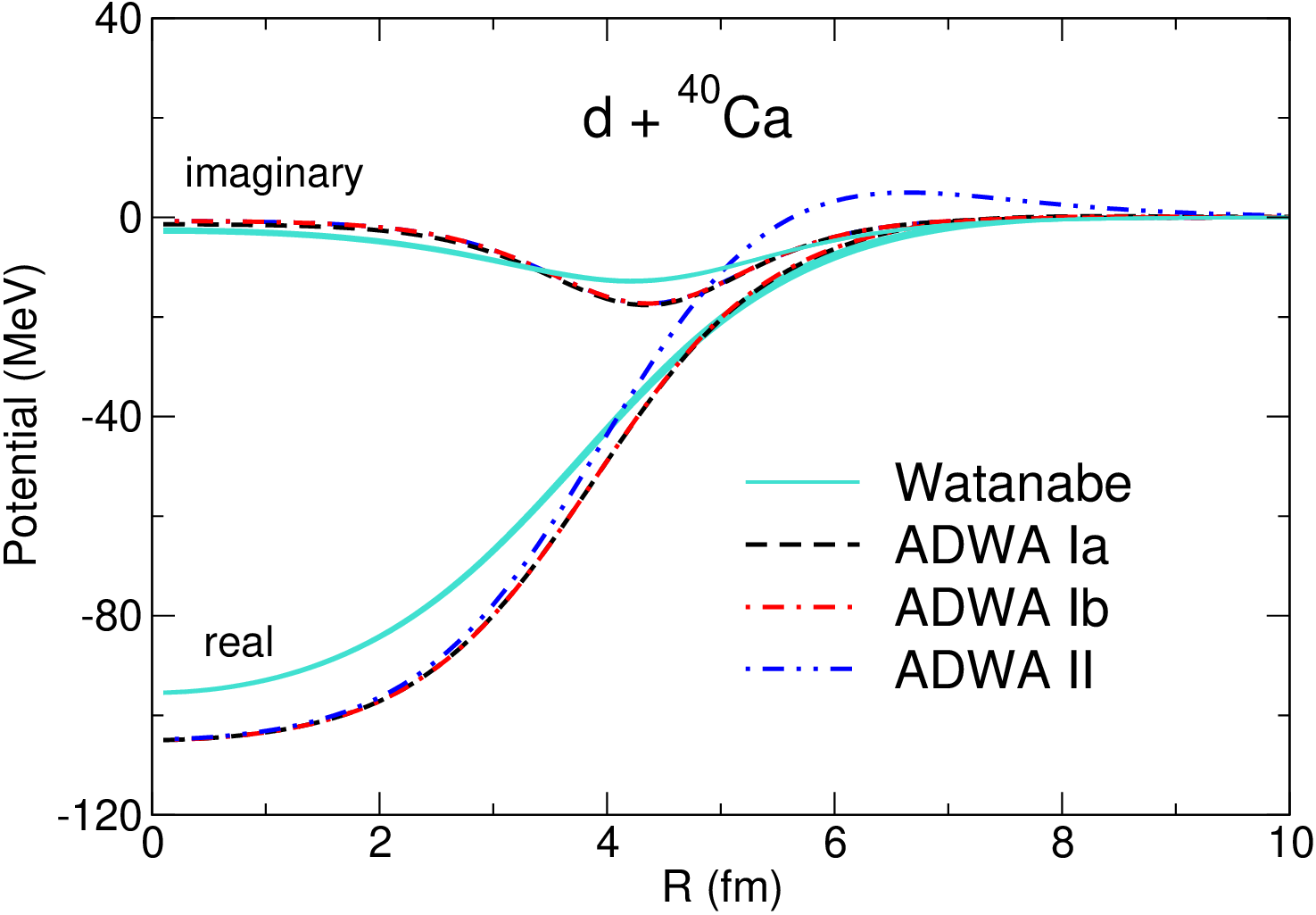}
}
\caption{The adiabatic  $d$-$^{40}$Ca potentials obtained in models Ia, Ib and II (dashed, dot-dashed and dot-dot-dashed lines) in comparison to the $d$-$^{40}$Ca potentials from the Watanabe model (thick band).}
\label{fig:adpots}
\end{figure}

The adiabatic potential in the deuteron channel has been calculated in  all three models, Ia, Ib and II, using the Hulth\'en  deuteron wave function $\phi_0(r)$ \cite{Hulthen}. The spin-orbit potential  both in the deuteron and proton channel were neglected. A proper treatment of the spin-orbit interaction in the ADWA requires calculating new tensor terms \cite{Joh15} for which no codes have been developed so far.
 Explicit calculations within distorted-wave-Born-approximation, performed here as a test   for $^{40}$Ca($d,p)^{41}$Ca at $E_d = 10$ and 20 MeV, show that the change in the cross section at the first maxima (the region used to determine spectroscopic factors) due to neglecting spin-orbit interaction in the entrance and/or exit channels are smaller than 4$\%$. The effective adiabatic potentials $U^{\rm eff}_{dA}$ are shown in Fig. \ref{fig:adpots}. One can see that their imaginary parts are almost identical in all three models. The difference between the real parts of $U^{\rm eff}_{dA}$,  obtained in Ia and Ib, is very small. The model II real potential is noticeably different. To understand where the difference comes from,  an approximation $r \gg R$ was made in all equation of section 4, as   is usually done in the Johnson-Soper model \cite{JS}, together with the assumption of an infinite mass $A$. This approximations works well for all standard values of diffuseness and should work even better for $a_{\rho} = 0.9$ fm. It was found that in this approximation, the Ia and Ib potentials are just the sums of the effective nucleon optical potentials $U^{\rm eff}_{NA}(R)$ while all the corrections due to these terms   are very small. In the case of model II, the only significant contribution, that supplements  the sum of nucleon optical potential $U^{\rm eff}_{nA}(R)+ U^{\rm eff}_{pA}(R)$,  comes from $B(R)$, which in the Johnson-Soper approximation reduces to $\rho(R)/(1-\rho(R)) \la T_{np} \ra_V$, where $\la T_{np} \ra_V = \la \phi_0 | V_{np}T_{np}| \phi_0\ra/ \la \phi_0 | V_{np} | \phi_0\ra$ is the kinetic energy of the $n$-$p$ pair averaged over the short-range of their interaction. The same quantity  is present in local-equivalent potentials in the nonlocal $(d,p)$ model \cite{Tim13a}. It is strongly  dependent on the choice of the deuteron model \cite{Bai16,Bai17}, inducing an enhanced sensitivity of the $(d,p)$ cross sections to the high $n$-$p$ momenta,   typical for short range $n$-$p$ separation. 

To calculate the $^{40}$Ca$(d,p)^{41}$Ca cross sections, the overlap integral between $^{41}$Ca and $^{40}$Ca ground states is needed. It  was taken from the nonlocal dispersive optical model (NLDOM) \cite{Mah14}. The updated parameters of this model are given in \cite{Wal16} where it was also shown that this overlap function can be approximated, with good accuracy, by the product of a single-particle wave function obtained for a  Wood-Saxon potential well (with $r_0$ = 1.252   fm, diffuseness $a = 0.718$ fm and spin-orbit depth $V_{s.o.} = 6.25$ MeV) and the square root of the spectroscopic factor $S = 0.73$. The same overlap function was used in all the calculations of this paper.

The deuteron distorted waves, generated by the ADWA potentials, were multiplied by the Perey factors of the corresponding models externally and then read back in by the TWOFNR code. The multiplication of deuteron distorted waves by the Perey factor reduces the $(d,p)$ cross sections by approximately 8$\%$ without noticeable changes in the shape of their angular distributions for all three models. The proton distorted wave in the exit channel were also multiplified by the nucleon Perey factor. This reduces the transfer cross sections further down by about 5$\%$. The  $^{40}$Ca+p  rather than $^{41}$Ca+p potential was used in the exit channel to remove the remnant term, as discussed in \cite{Tim99}. 

\begin{figure}[h!]
\centering
\includegraphics[scale=0.4]{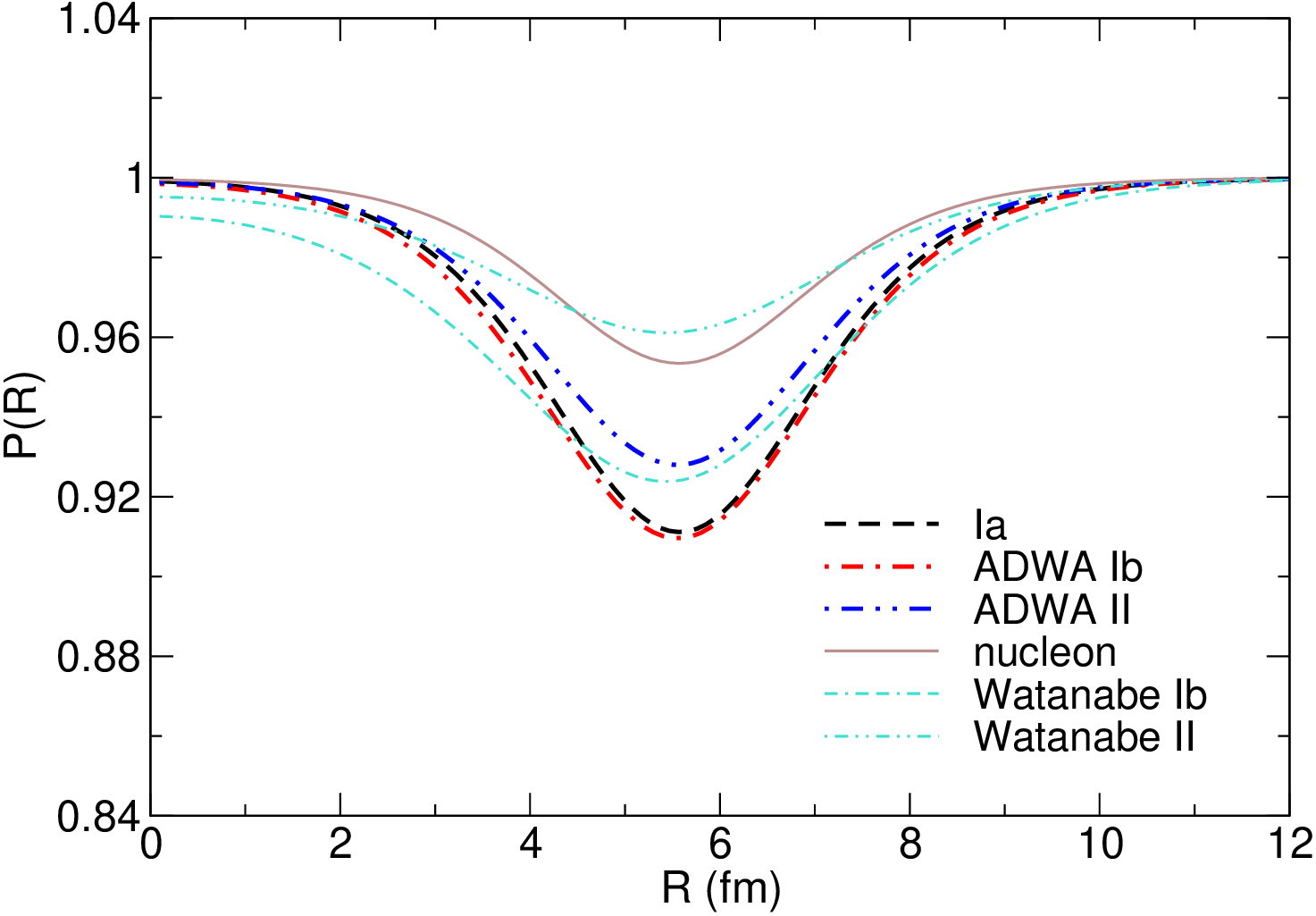}
\caption{The Perey factors in the $d-^{40}$Ca channel, calculated in the ADWA and the Watanabe models, in comparison with the $p-^{40}$Ca Perey factor (thin line) given be Eq. (\ref{pvm}). For the Ia model, the ADWA and Watanabe Perey factors are identical. }
\label{fig:Pereyfactors}
\end{figure}

The Perey factors $P(R)$ are shown in Fig. \ref{fig:Pereyfactors}.
They are almost identical for the Ia and Ib models, being   approximately equal to the square of the nucleon Perey factor $P_N$ from Eq. (\ref{pvm}), also shown in this figure.  The reason behind this near equality is the same as the one, discussed above, that causes the adiabatic Ia and Ib potentials to be very similar. Due to the short range nature of $\phi_1(r)$ the model Ib value of $S(R)$  is approximately equal to $\frac{1}{2}[\nabla_R F_n(R)+\nabla_RF_p(R)]$. Furthermore, if $F_n=F_p \equiv F$  then   equation (\ref{PIb}), of which $P(R)$ is the solution, in the limit of infinitely large mass $A$ transforms into $P^{\prime}/P = 2 m/\hbar^2 \nabla F$, where $m$ is the nucleon mass. This equation is the same as   Eq. (\ref{pereyeq}) that determines $P_n$ or $P_p$, apart from a factor of two, which causes the solution for $P(R)$ to be the square of $P_N$. The same $P(R)$ value is given by the model Ia: $P(R) = P_n(R)P_p(R) \approx P_N^2(R)$. This similarity gradually disappears with decreasing  diffuseness of $\rho(r)$  because  the validity of the  Johnson-Soper $r \gg R$  estimate  for $S(R)$ is getting lost. As for the Perey factor for model II, it comes from a differential equation with ${\tilde S}(R)$ that does not reduce to $S(R)$ and, therefore, it cannot be equal to the $P(R)$ from model I. Figure \ref{fig:Pereyfactors} shows that it is smaller. 

\begin{figure}[h!]
\centering
\includegraphics[scale=0.4]{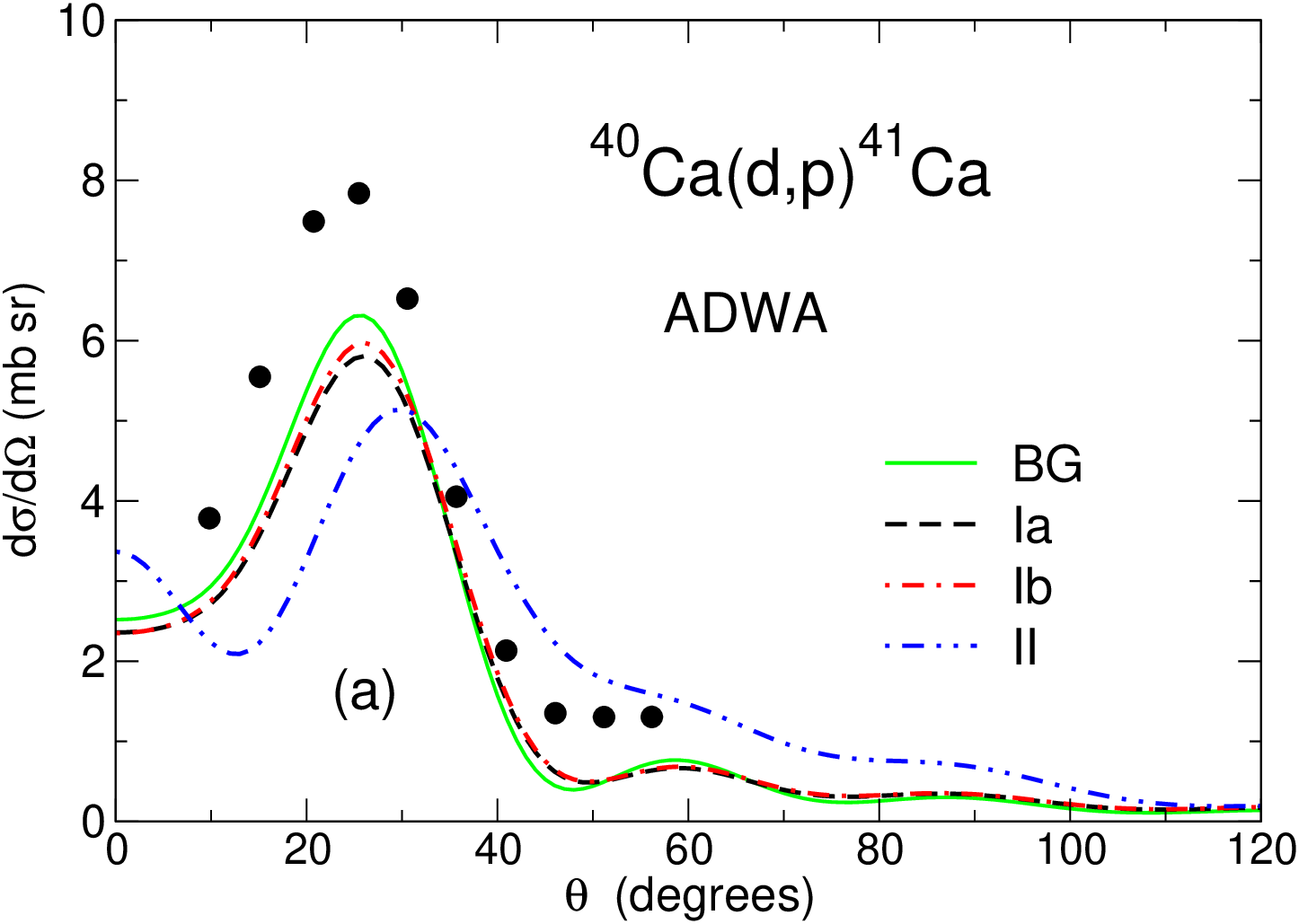}
\includegraphics[scale=0.4]{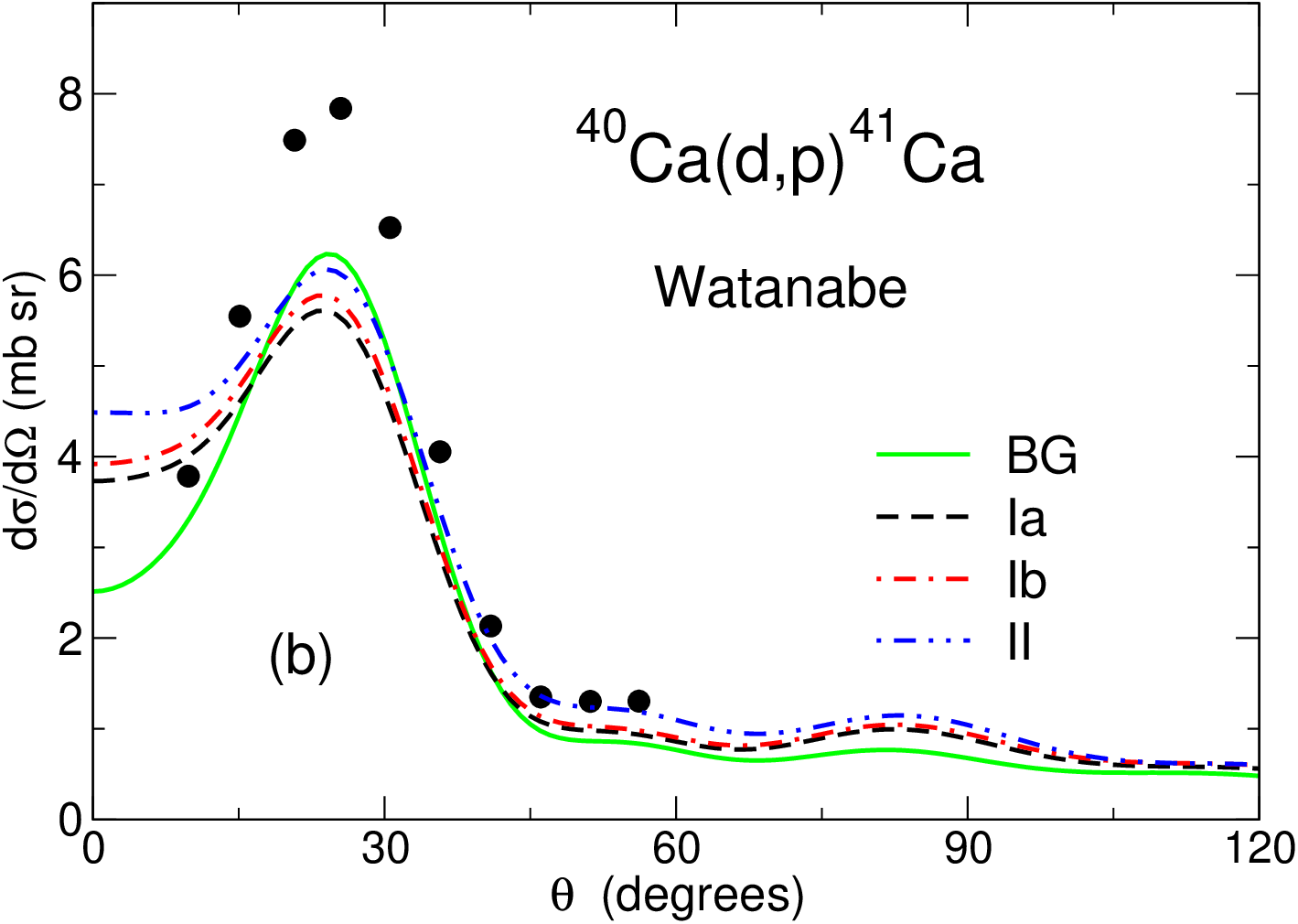}
\caption{The $^{40}$Ca$(d,p)^{41}$Ca cross sections at $E_d = 20 $ MeV calculated using   velocity-dependent optical nucleon potentials  in three different models of the ADWA ($a$) and Watanabe ($b$) approach in comparison to the standard local calculations with Becchetti-Greenless optical potentials.
}
\label{fig:xsecs}
\end{figure}

The calculated $^{40}$Ca$(d,p)^{41}$Ca cross sections are shown in Fig. \ref{fig:xsecs}a. One can see that models Ia and   Ib give   similar predictions for these cross sections, which are very close to those calculated in the Johnson-Tandy local model with the original Becchetti-Greenlees potential. However,  model II gives a significantly different result, both in   shape and magnitude, which is the consequence of the additional term in Eq. (\ref{VdAR}),  $B(R)/(1-{\cal R}(R))$, determined by high $n$-$p$ momenta in deuteron through $\la T_{np} \ra_V$. It has been pointed out in \cite{Gom18} that nonlocal non-diagonal potentials, that couple the first Weinberg component $\chi_0$ to all the other ones,   also strongly depend on the deuteron model choice through their different high $n$-$p$ momentum content and that a large number of Weinberg states is needed to overcome this model-dependence. It was also shown in \cite{Gom18} that an alternative expansion of the three-body wave function $\Phi(\ve{R},{r})$ over the CDCC basis does not generate significant $n$-$p$ momentum dependence, which has also been confirmed by rigorous Faddeev calculations in \cite{Del18}.

The first term of the CDCC expansion is given by the Watanabe folding model and for velocity-dependent nucleon optical potentials the Watanabe potentials can be obtained by  replacing $\phi_1(\ve{r})$  with $\phi_0(\ve{r})$ everywhere in section 4. The Watanabe potentials, obtained in models Ia, Ib and II, are very similar (see Fig. \ref{fig:adpots}) because   $B(R)$ is now determined by the average $n$-$p$ kinetic energy over all the coordinate space, which is much lower than $\la T_{np}\ra_V$. The Perey factors calculated in models Ib and II are smaller and wider. Their influence on $(d,p)$ sections is similar to the ADWA case. The Watanabe
cross sections are shown in Fig. \ref{fig:xsecs}b. There is no significant difference between all three models. Some difference between the Watanabe cross sections obtained with and without velocity-dependent forces,using the original Becchetti-Greenlees potential only, is seen at very small angles.

The calculated $^{40}$Ca($d,p)^{41}$Ca cross sections are compared in Figs. \ref{fig:xsecs}a and \ref{fig:xsecs}b to the experimental data from   \cite{Eck90}. Both the  ADWA and the Watanabe approaches predict 30$\%$ smaller cross sections as compared to the experimental data, thus  suggesting that the spectroscopic factor should be larger than the NLDOM  value of 0.73 used in these calculations. However, it was shown in \cite{Ngu11} that using a different optical potential, such as local dispersive-model optical potential from \cite{Mue11}, results in smaller spectroscopic factors. 
Also, a contribution from other reaction mechanisms could be responsible for disagreement between the predicted and measured cross sections.


\section{Summary and conclusions}

In this paper, velocity-dependent optical potentials that have a form  consistent with a spatially-variable  mass of a nucleon interacting with nucleons of the target, have been considered within the context of the three-body $A + n + p$ problem. For such a class of  potentials an exact local-equivalent two-body model exist. As a consequence, there are two ways to formulate the three-body problem and to write down the  corresponding Schr\"odinger equation. The first one  is based on nucleon  potentials with  only first derivatives over the $p-A$ and $n-A$ coordinates, while the second one  is based on original velocity-dependent potentials that contain  both first and second derivatives on these coordinates. There is no obvious connection between these two representations at a formal level.

The main reason to look into the velocity-dependence of optical potentials was to study the manifestation of nonlocal effects in $(d,p)$ reactions.  Velocity-dependence is a particular form of nonlocality that can be derived  for a general nonlocal potential by considering only a linear on kinetic energy term in its Taylor expansion  over a variable that determines the nonlocality range. A three-body Schr\"odinger equation with velocity-dependent potentials could be an easier task to solve than the one that involves nonlocal potentials of a general nature. It has been solved in the adiabatic approximation, which accounts for deuteron breakup and provides the wave functions at small $n$-$p$ separations, relevant to the $(d,p)$ problem. This results in distorted waves $\chi(\ve{R})$   determined from a two-body differential equation  with an adiabatic potential that could be constructed in  three different ways.

The local-equivalents of velocity-dependent potentials, originally proposed in \cite{Jag12a,Zur13}, have a strongly repulsive surface feature,  originating due to a  very small diffuseness of velocity-dependent term, which is impossible to understand. Increasing  the diffuseness, as suggested in the present paper, removes such an  unphysical behaviour. 
Application of the velocity-dependent force, proposed in this work,  together with the local part  fixed by  Becchetti-Greenlees systematics,
to the $^{40}$Ca$(d,p)^{41}$Ca reaction revealed a large difference   between the calculations based on three-body Schr\"odinger equation with first derivatives only (model I) and the calculations that include both the first and the second derivatives (model II). This difference is easily explained by an additional term in model II associated with the high $n$-$p$ momenta in the $n$-$p$ kinetic energy in deuteron averaged over the short range of the $n$-$p$ interaction, similar to what has been observed in  a more general nonlocal problem  discussed in \cite{Bai16}. Based on recent studies in \cite{Gom18,Del18} it is reasonable to expect that  this difference will most likely decrease when the three-body Schr\"odinger equation is solved beyond the adiabatic approximation. Indeed, the $(d,p)$ calculations, performed here in the Watanabe model (which comes from the lowest term of the CDCC expansion), do  not show any difference between models I and II.

The energy of the $^{40}$Ca($d,p)^{41}$Ca reaction has been chosen here because of the availability of experimental data both for the reaction and for the elastic scattering in the entrance and exit channels. However, comparison between the theory and experiment suggests that the spectroscopic factor of 0.73, rooted in the NLDOM  (a theory that forges a link between the nuclear structure and nuclear reactions \cite{Mah14}) and  used in the calculations, is small. This contradicts a previous study at a lower deuteron energy \cite{Wal16} with the same overlap integral and could be explained by a need for a different $p-^{40}$Ca optical potential. It is also possible that the reaction mechanism at this energy is not sufficiently understood. More theoretical work as well as more $^{40}$Ca$(d,p)^{41}$Ca reaction measurements around 20 MeV  are needed to understand the spectroscopic factor  of $^{41}$Ca as determined from transfer reactions. 

Finally, to make a judgement  about the importance of a velocity-dependent force in $(d,p)$ reactions, this force should be, first of all,  unambiguously extracted  from an independent source. Fitting this force to elastic scattering data will not provide a unique solution. Fixing this force from nuclear structure calculations while fitting the local part of the optical potential only could be an optimal way to go forward. More research is needed in this direction. A more realistic treatment of optical potentials will help to reduce the uncertainties of spectroscopic information extracted from $(d,p)$ reactions.

\section*{Acknowledgements}
This work was supported by the United Kingdom Science and Technology Facilities Council (STFC) under Grants No. ST/L005743/1 and ST/P005314/1.

\end{document}